\pdfoutput=1

\documentclass[sigconf, screen, nonacm]{acmart}

\usepackage{enumitem}
\setlist[itemize]{topsep=4pt}

\usepackage{xspace}
\usepackage{mathrsfs}%
\usepackage{tikz}
\usepackage{makecell}%
\usepackage{tabularx}%
\usepackage{colortbl}%
\definecolor{oursbg}{RGB}{226,236,248}

\newcommand{\us}{UnitSight\xspace}
\newcommand{\fs}{ForeSight\xspace}

\newcommand{\Czero}{\ensuremath{\mathscr{C}_0}\xspace}
\newcommand{\Cone}{\ensuremath{\mathscr{C}_1}\xspace}

\newcommand{\step}[1]{\tikz[baseline=(s.base)]\node[circle,fill=black,text=white,inner sep=0pt,minimum size=1.1em,font=\scriptsize\bfseries](s){#1};}
\definecolor{usRed}{HTML}{E90000}
\newcommand{\ustep}[1]{\tikz[baseline=(s.base)]\node[circle,fill=usRed,text=white,inner sep=0pt,minimum size=1.1em,font=\scriptsize\bfseries](s){#1};}

\newcommand{\observation}[2]{\par\vspace{2pt}\noindent\hfill{\setlength{\fboxsep}{3.5pt}\setlength{\fboxrule}{0.5pt}%
\fbox{\parbox{\dimexpr0.97\columnwidth-2\fboxsep-2\fboxrule\relax}{\textbf{Observation\ifx&#1&\else~#1\fi:}\enspace#2}}}\hfill\null
\par\vspace{2pt}}

\begin{document}

\title{Exploiting What NVIDIA Profilers Hide: Fast and Accurate Analytical GPU Performance Modeling}

\author{Ziwen Wang}
\affiliation{%
  \institution{The Hong Kong University of Science and Technology (Guangzhou)}
  \city{Guangzhou}
  \country{China}}
\email{zwang578@connect.hkust-gz.edu.cn}

\author{Qi Luo}
\affiliation{%
  \institution{The Hong Kong University of Science and Technology (Guangzhou)}
  \city{Guangzhou}
  \country{China}}
\email{qluo615@connect.hkust-gz.edu.cn}

\author{Kunlin Li}
\affiliation{%
  \institution{The Hong Kong University of Science and Technology (Guangzhou)}
  \city{Guangzhou}
  \country{China}}
\email{kli082@connect.hkust-gz.edu.cn}

\author{Changxi Liu}
\authornote{Corresponding authors.}
\affiliation{%
  \institution{The Hong Kong University of Science and Technology (Guangzhou)}
  \city{Guangzhou}
  \country{China}}
\email{changxiliu@hkust-gz.edu.cn}

\author{Dongsheng Wang}
\affiliation{%
  \institution{Tsinghua University}
  \city{Beijing}
  \country{China}}
\email{wds@tsinghua.edu.cn}

\author{Yun Chen}
\authornotemark[1]
\affiliation{%
  \institution{The Hong Kong University of Science and Technology (Guangzhou)}
  \city{Guangzhou}
  \country{China}}
\email{yunchen@hkust-gz.edu.cn}

\begin{abstract}
GPUs have become essential to many fields, including AI and scientific computing.
As GPUs grow more complex, the design space of their hardware configurations becomes increasingly large, yet cycle-level simulators remain too slow to explore it.
Analytical models make this exploration much faster.
However, existing GPU analytical models fall short of accurately predicting performance across hardware configurations, because they re-estimate or scale every stall of a warp on the target configuration.
We observe that fixed-latency stalls stay the same across configurations, whereas queueing stalls vary with the pressure on each hardware unit.
Based on this observation, we propose \fs, a queueing analytical model that reuses the fixed-latency stalls measured on GPU hardware and recomputes the queueing stalls of each hardware unit from its pressure.
To provide \fs with the counter values of each hardware unit, we build \us, a profiler that recovers these values, which NVIDIA's official profilers hide by aggregating across units.
Our evaluation on hardware configurations of nine NVIDIA GPUs shows that \fs predicts kernel execution time with a mean absolute percentage error (MAPE) of 8.3\%, compared with 14.1\% for the best baseline.
\fs also predicts the end-to-end execution time of LLM inference on vLLM with a MAPE of 2.0\%.
It runs 49$\times$ faster than the cycle-level simulator Accel-Sim and faster than interval-based models such as GCoM.
With the aid of \fs, we explore configurations derived from the H20 GPU for LLM inference and find one with 1.78$\times$ its performance/cost.

\end{abstract}

\maketitle

\section{Introduction}
\label{sec:intro}

GPUs have become the workhorse of AI and scientific computing, powering tasks from training large models to serving inference for massive user bases.
These workloads place different demands on hardware resources.
For example, training and prefill are compute-bound, whereas decode is memory-bandwidth-bound.
To meet these diverse demands, vendors derive several GPUs from the same microarchitecture.
These GPUs share the same streaming multiprocessor (SM) design but differ in their hardware configurations: the number of SMs, the capacity and bandwidth of the L2 cache and DRAM, and the clock frequency.
Architects must therefore predict the performance of workloads across many candidate hardware configurations to select the configuration of each GPU, a process known as design space exploration (DSE).

Cycle-level simulators \cite{khairy2020accelsim,liu2023photon} are the most widely used prediction tools in DSE.
They model each hardware unit of a GPU in microarchitectural detail and simulate the behavior and interactions of these units cycle by cycle.
Simulators can evaluate arbitrary hardware configurations, but they are extremely slow.
For faster prediction, researchers have proposed performance models, including data-driven methods \cite{lee2025neusight,zhang2026pipeweave} and analytical models \cite{huang2014gpumech,wang2020mdm,lee2022gcom,cao2025amali,wang2018gpgpu}.
Rather than simulating hardware behavior cycle by cycle, they directly estimate workload execution time.
Data-driven methods train machine learning models on execution times measured across multiple GPUs to learn how workload features and hardware configurations relate to execution time, but these models generalize poorly to workloads and GPUs outside their training distribution.
Analytical models capture the first-order factors of GPU execution in mathematical equations.
Among them, interval-based analytical models \cite{huang2014gpumech,wang2020mdm,lee2022gcom,cao2025amali,cha2025gcscaler} compute the execution time of a representative warp as the sum of its issue and stall cycles on the target configuration, and then estimate how concurrent warps hide stalls and contend for hardware units.
Scaling-based analytical models \cite{yu2021habitat,nath2015crisp} first measure workload execution time on GPU hardware and then scale it to the target hardware configuration by the ratios of parameters such as bandwidth and frequency.

However, we observe that not all stalls of a warp change when the hardware configuration changes.
Fixed-latency stalls, in which an instruction waits a fixed number of cycles for a pipeline result, stay the same.
Queueing stalls, in which warps or memory requests wait at a busy hardware unit, vary because the configuration changes the pressure on that unit.
Interval-based analytical models re-estimate all stalls on the target configuration from formulas, and derive memory queueing stalls from the bandwidth of the whole GPU rather than from the pressure on each unit.
Scaling-based analytical models instead scale both types of stalls by the same parameter ratios, so they change fixed-latency stalls and derive queueing stalls from these ratios rather than from pressure.

Based on this observation, we propose \fs, a queueing analytical model.
To predict the execution time of a workload on a target configuration, \fs reuses the fixed-latency stalls measured on GPU hardware and recomputes the queueing stalls.
Because the pressure can differ across hardware units, \fs computes queueing stalls for each unit individually.
However, existing profilers (NCU \cite{ncu} and CUPTI \cite{cupti}) cannot provide the pressure on each unit because they report only aggregated values across units.
We find that these profilers actually collect the counter values of every hardware unit but aggregate these values before reporting them, which hides this pressure.
Based on this finding, we build \us, which recovers these values before the profilers aggregate them and provides them to \fs.
With these per-unit values, \fs replays the dispatch of CTAs to the SMs of the target configuration and, at each moment of the replay, solves the queueing at the pipelines of each SM and at each L2 slice and DRAM channel.
Beyond providing per-unit values to \fs, \us, as a profiler, can also help developers analyze the performance bottlenecks of their kernels.
Its per-unit values reveal what aggregated metrics hide, such as load imbalance across SMs.
It also allows a kernel to read performance counters with PTX or SASS instructions, so developers can profile any code region of the kernel rather than only the whole kernel.

To evaluate the accuracy of \fs in DSE, we construct diverse hardware configurations across nine NVIDIA GPUs by varying the SM count, the SM frequency, and memory resources, and compare its predictions with the real execution times on these GPUs.
\fs predicts execution time with a mean absolute percentage error (MAPE) of 8.3\%, compared with 14.1\% for the best baseline.
For end-to-end inference of Llama-3.1-8B on vLLM, \fs predicts the total execution time with a MAPE of 2.0\%, whereas the baselines range from 14.3\% to 137\%.
\fs runs 49$\times$ faster than the cycle-level simulator Accel-Sim \cite{khairy2020accelsim} and 1.6$\times$ and 4.2$\times$ faster than the interval-based models GCoM \cite{lee2022gcom} and AMALI \cite{cao2025amali}.
Finally, as a case study, we use \fs to predict LLM inference on hardware configurations derived from the H20 and find one whose performance/cost is 1.78$\times$ that of the H20.

This paper makes the following contributions:
\begin{itemize}
    \item We characterize how the stalls of a warp change across hardware configurations, finding that fixed-latency stalls stay the same while queueing stalls vary with the pressure on each hardware unit.
    \item We build \us, the first profiler for NVIDIA GPUs since Turing that recovers the counter values of each hardware unit, which NVIDIA's official profilers hide by aggregating across units.
    \item We propose \fs, a fast and accurate queueing analytical model that predicts GPU performance across hardware configurations by reusing the fixed-latency stalls measured on GPU hardware and recomputing the queueing stalls of each hardware unit.
    \item We evaluate \fs on hardware configurations of nine NVIDIA GPUs, where it predicts kernel and end-to-end LLM inference time with MAPEs of 8.3\% and 2.0\%. It runs 49$\times$ faster than Accel-Sim and faster than interval-based models. In a DSE case study, it finds a configuration derived from the H20 with 1.78$\times$ its performance/cost.
\end{itemize}

\section{Background and Motivation}
\label{sec:background}

\subsection{GPU Architecture and Profiling}\label{sec:bg-arch}

GPU compute resources are organized hierarchically into Graphics Processing Clusters (GPCs), Streaming Multiprocessors (SMs), and SM sub-partitions (SMSPs). A kernel is divided into cooperative thread arrays (CTAs) dispatched to SMs, while warps serve as the basic scheduling units within each SMSP. Each SMSP contains a warp scheduler and multiple execution pipelines, such as the ALU, FMA, and tensor pipelines. Every cycle, the warp scheduler selects a ready warp and issues an instruction to a pipeline. When the warp scheduler cannot issue an instruction from a warp, the warp stalls, for instance when waiting for an operand from a prior instruction or for data from memory. A memory request first accesses the per-SM L1 cache, which serves as both data cache and shared memory. Requests missing in the L1 cache traverse an on-chip crossbar to reach the L2 cache, which consists of multiple L2 slices. Requests missing in the L2 cache proceed to off-chip DRAM, which consists of multiple DRAM channels. Each request goes to the L2 slice and DRAM channel that its address maps to. We refer to warp schedulers, pipelines, L1 caches, L2 slices, and DRAM channels as hardware units. Each hardware unit has performance monitoring (PM) counters, which record events such as instruction counts, request counts, and stall cycles. NVIDIA's profilers, Nsight Compute (NCU)~\cite{ncu} and CUPTI~\cite{cupti}, configure these counters and compute the metrics that the user specifies. Each metric aggregates the counts of all units of the same type, such as their sum or average.

\subsection{Motivation}\label{sec:motivation}

Existing analytical models lose accuracy when the hardware configuration changes. We restrict 50 kernels from LLM inference and microbenchmarks to fewer SMs of an H100 PCIe GPU with green contexts~\cite{nvidia2026greenctx}. With all 114 SMs, Habitat~\cite{yu2021habitat}, a scaling-based model, predicts their execution times with a MAPE of 1.1\%, but with 16 SMs its MAPE reaches 42\%. The MAPE of GCoM~\cite{lee2022gcom}, an interval-based model, grows from 43\% to 82\%. To understand why, we examine how the stalls of these kernels change with the configuration.

\begin{figure}[t]
  \centering
  \includegraphics[width=\columnwidth]{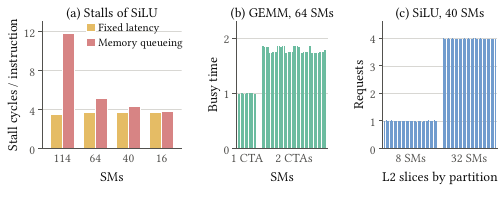}
  \caption{\textbf{Stalls, SM busy time, and L2 requests on H100 PCIe.}}
  \label{fig:motiv}
\end{figure}

\begin{figure*}[t]
  \centering
  \includegraphics[width=\textwidth]{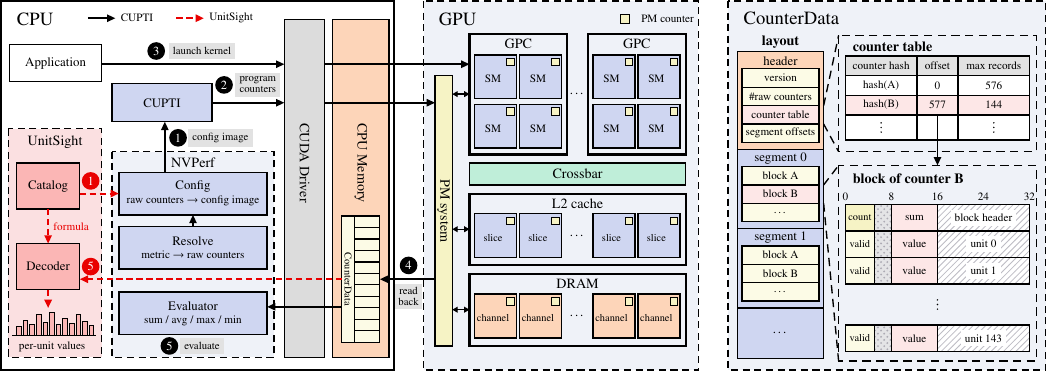}
  \caption{\textbf{UnitSight overview.} CUPTI's profiling flow is in black, and the parts UnitSight replaces are in red. The right panel shows the layout of CounterData.}
  \label{fig:unitsight}
\end{figure*}

These stalls arise from two sources: pipeline fixed latencies and queueing at hardware units. A fixed-latency stall occurs when a dependent instruction waits a fixed number of cycles for an earlier instruction to finish in a pipeline, so it does not change with the configuration. A queueing stall occurs when more warps or memory requests reach a hardware unit than it can serve at once: warps queue at the pipelines inside an SM, and memory requests queue at L2 slices and DRAM channels. When the configuration changes, the number of warps and requests using each hardware unit changes, so queueing stalls change as well. We verify this distinction on the SiLU kernel by measuring its stall cycles per instruction with 114 and 16 SMs. Fixed-latency stalls remain nearly unchanged (3.5 to 3.7 cycles per instruction), whereas memory queueing stalls drop from 11.8 to 3.8 cycles per instruction (Figure~\ref{fig:motiv}a). Interval-based models derive both types of stalls on the target configuration from formulas and latency tables, so the errors of these estimates enter the prediction. For example, in the experiment above, GCoM predicts 0.89~ms for the SiLU kernel with 16 SMs, whereas the kernel actually takes 0.41~ms. Scaling-based models multiply the measured time by a single factor built from hardware ratios, so the fixed-latency stalls are scaled together with the queueing stalls instead of staying constant, and the queueing stalls are not recomputed from the new pressure. In the same experiment, Habitat predicts 0.27~ms for the SiLU kernel with 16 SMs, 34\% below the measured 0.41~ms.

Moreover, the queueing at each hardware unit grows with the pressure on that unit, and this pressure can differ across units and change with the configuration. Across SMs, the pressure differs when the CTAs of a kernel cannot be dispatched evenly to the SMs, and some SMs then receive more CTAs and more work than others. This load imbalance across SMs changes with the number of SMs. Across L2 slices and DRAM channels, the pressure differs when the address hash maps the requests of a workload unevenly to these units. This happens when the requests fall in a small address range. In addition, on some GPUs, the L2 slices and DRAM channels are divided into two partitions, and each SM sends its requests only to the slices of its own partition~\cite{liu2026dgna}. When the two partitions hold different numbers of active SMs, their slices receive different pressure, and so can their DRAM channels.

\looseness=-1 We verify this by varying the SMs assigned to two kernels with green contexts on the H100 PCIe GPU. The FP8 GEMM kernel of the QKV projection in Llama-3.1-8B always launches 114 CTAs. With 64 SMs, these CTAs cannot be dispatched evenly, and 50 SMs stay busy 1.8$\times$ as long as the other 14 (Figure~\ref{fig:motiv}b). Similarly, for the SiLU kernel with 40 SMs, 8 in one partition and 32 in the other, 40 L2 slices receive 3.98$\times$ as many requests as the other 40 (Figure~\ref{fig:motiv}c). Therefore, the same workload can put very different pressure on individual hardware units under different configurations, and queueing stalls must be recomputed for each unit rather than from GPU-wide totals.

\observation{}{When the hardware configuration changes, fixed-latency stalls remain unchanged and can be reused, whereas queueing stalls change and must be recomputed from the pressure on each hardware unit.}

\fs (Section~\ref{sec:foresight}) builds on this observation and recomputes the queueing stalls of each hardware unit from its pressure. However, NVIDIA's profilers, NCU and CUPTI, report only metrics aggregated across units, which hide this pressure. We therefore build \us (Section~\ref{sec:unitsight}), which profiles each hardware unit individually and thus reveals this pressure.

\section{UnitSight}
\label{sec:unitsight}

This section presents UnitSight. Section~\ref{sec:us-cupti} analyzes CUPTI's profiling flow and the layout of CounterData, Section~\ref{sec:us-design} presents the design of UnitSight and its in-kernel PM reading mode, Section~\ref{sec:us-validation} validates its per-unit values, and Section~\ref{sec:us-case} uses a case study to show how it diagnoses a real LLM kernel.

\subsection{CUPTI Profiling Flow and CounterData}
\label{sec:us-cupti}

At the hardware level, an NVIDIA GPU tracks performance per unit: each SM, each L2 slice, and each DRAM channel has its own PM counters connected to the GPU's PM system (Figure~\ref{fig:unitsight}, middle). At the profiler level, on Volta and earlier GPUs, NVIDIA's previous profiler nvprof and CUPTI's previous Event API reported the values of each unit's PM counters~\cite{nvprof,cupti}. Starting with Turing, however, they no longer support counter collection and are replaced by the new profiler NCU and CUPTI's new Profiling API. NCU and the Profiling API report only aggregates of these per-unit values: the sum, average, maximum, or minimum across all SMs, all L2 slices, or all DRAM channels. To understand this gap, we study CUPTI's profiling flow (Figure~\ref{fig:unitsight}, left). First, the user specifies the metrics to profile, and NVPerf, a closed-source library, resolves them into raw counters, the hardware events counted by the PM counters (Resolve). From these raw counters, NVPerf builds the config image (Config), a configuration assigning each PM counter a raw counter, and passes it to CUPTI (\step{1}). CUPTI then writes this config image through the CUDA driver into the GPU's PM system (\step{2}), and the PM system sets each PM counter of every unit to its assigned raw counter. With the counters configured, the application launches the kernel through the CUDA driver onto the SMs (\step{3}), and every unit's PM counters record events during execution. After the kernel finishes, the PM system reads the resulting counter values back into CounterData, a buffer in CPU memory (\step{4}). Finally, NVPerf's Evaluator computes the requested metrics from CounterData and returns them to the user (\step{5}).

We find that after the read-back, CounterData does contain the counter value of every unit, but NVPerf's Evaluator aggregates these values into the metrics it returns, which hides the value of each unit. NCU uses the same NVPerf library to compute metrics, and therefore also reports only aggregates. We analyze the layout of CounterData (Figure~\ref{fig:unitsight}, right): it consists of a header followed by one segment per profiled kernel, each containing one block per raw counter. The header holds the format version of CounterData (version), the number of raw counters resolved by NVPerf (\#raw counters), the counter table, and the offset of each segment (segment offsets). Each row of the counter table describes a raw counter by the 64-bit hash NVPerf computes to identify it (counter hash), the offset of its block within a segment (offset), and the number of per-unit records reserved in the block (max records). A block starts with a block header holding the number of records with data (count) and the sum of their values (sum), followed by the reserved records. Each record has a flag telling whether it holds data (valid) and the counter value of one unit (value). The value field, however, does not store the counter value directly. We find that the counter value of a unit is the XOR of the content of its value field after CUPTI initializes CounterData before profiling and after the read-back, and the count, sum, and valid fields are also computed by XOR.

With this layout, we can read the counter value of every unit for each raw counter in each kernel. To read raw counter B on unit $i$ in kernel $k$, we locate the segment of kernel $k$ from the segment offsets in the header, find B's row in the counter table by its counter hash, follow its offset (577) to B's block in that segment, and XOR the value field of unit $i$'s record after initialization and after the read-back.

\subsection{UnitSight Design}
\label{sec:us-design}

Based on these findings, we build UnitSight, the first hardware-counter-based profiler for NVIDIA GPUs since Turing that can profile each hardware unit individually. As shown in red in Figure~\ref{fig:unitsight}, UnitSight keeps CUPTI's counter collection (\step{2}--\step{4}), but replaces NVPerf's Resolve with a catalog that maps metrics to raw counters (Catalog, \ustep{1}) and NVPerf's Evaluator with a decoder reading the per-unit values directly from CounterData (Decoder, \ustep{5}). Beyond the flow in Figure~\ref{fig:unitsight}, UnitSight can also read an SM's PM counters inside the kernel to obtain the counter values of a code region, a capability that CUPTI and NCU do not support.

In CUPTI, the user can only specify metrics, and NVPerf's Resolve decides which raw counters to configure. To allow the user to configure raw counters directly, UnitSight builds the Catalog. The Catalog is a table listing all raw counters of a GPU chip and the raw counters used by each metric, and we build one for every chip UnitSight supports. For example, the Catalog for the H100 GPU covers 2,785 raw counters and about 236,000 metrics. The user selects raw counters from the Catalog, and the Catalog passes them to NVPerf's Config to build the config image. The user can also specify metrics as in CUPTI, and the Catalog converts each metric into its raw counters before passing them to Config.

\looseness=-1 UnitSight then profiles the kernel with CUPTI as usual: CUPTI writes this config image into the GPU's PM system, the application launches the kernel, and the PM system reads the counter values back into CounterData. Instead of NVPerf's Evaluator, UnitSight passes CounterData to the Decoder, which extracts the value of each raw counter on each unit in each kernel, using the method described in Section~\ref{sec:us-cupti}. To turn these values into the metrics the user specifies, the Decoder needs the expression computing each metric from its raw counters, which we call the metric's \emph{formula}. For instance, the formula of the L2-slice hit-rate metric, \texttt{lts\_\_t\_\allowbreak sector\_\allowbreak hit\_rate.pct}, is $100 \times N_{\mathrm{hit}} / N_{\mathrm{all}}$, where $N_{\mathrm{hit}}$ and $N_{\mathrm{all}}$ are the raw counters of hit sectors and of all accessed sectors in the slice. We obtain each metric's formula by filling CounterData with test values and fitting it to the metrics computed by NVPerf's Evaluator, and store the formula in the Catalog. The Decoder then applies the formulas from the Catalog to compute the metrics of each unit from its raw counters. CUPTI and NCU, in contrast, report only aggregates.

Like CUPTI and NCU, UnitSight provides two profiling modes: range profiling and PM sampling. Range profiling accumulates the counter values over a range and gives one value per range. By default, each kernel is a range, and the user can also mark an interval containing several kernels as one range. PM sampling reads the PM counters at a fixed time interval during execution, giving a time series of values, so changes within a kernel become visible. With the Catalog and the Decoder, UnitSight gives one value per unit for each range and one value per unit for each sample.

\begin{figure}[t]
  \centering
  \includegraphics[width=\columnwidth]{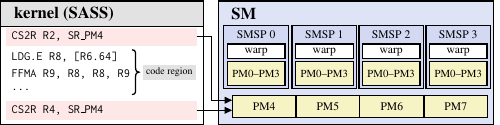}
  \caption{\textbf{In-kernel PM reading.} The kernel reads \texttt{SR\_PM4} before and after a code region. Each SMSP has its own PM0--PM3, and the SM shares PM4--PM7.}
  \label{fig:inkernel}
\end{figure}

Beyond these two modes, UnitSight also allows a kernel to read the PM counters of its own SM during execution, which we call \emph{in-kernel PM reading}. It is a combination of instrumentation-based and hardware-counter-based profiling. We find that PTX and SASS, the virtual and native assembly languages of NVIDIA GPUs, both provide special registers for PM counters (\texttt{\%pm0}--\texttt{\%pm7} in PTX, \texttt{SR\_PM0}--\texttt{SR\_PM7} in SASS), and that these registers count the raw counters configured through Config. Because UnitSight allows the user to configure raw counters directly, the user can decide which raw counter each register counts. We further observe that PM0--PM3 accept only SMSP-level raw counters, and a warp reading PM0--PM3 gets the counter value of its own SMSP. PM4--PM7 accept only SM-level raw counters, and a warp reading PM4--PM7 gets the counter value of the whole SM. Figure~\ref{fig:inkernel} shows an example of profiling a code region of a kernel by reading a PM register in SASS: the kernel reads \texttt{SR\_PM4} with the instruction \texttt{CS2R} before and after the code region, and the difference is the value of PM4's raw counter on the SM during the code region. With in-kernel PM reading, UnitSight can locate a kernel's bottleneck to a specific code region and explain it with the hardware counters of that region, which neither the other profiling modes nor instrumentation can do. For example, Section~\ref{sec:us-case} locates the bottleneck of FlashAttention-3 decode using this mode.

\begin{figure*}[t]
  \centering
  \includegraphics[width=\textwidth]{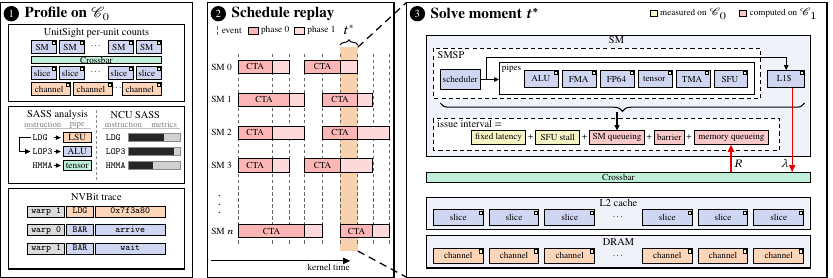}
  \caption{\textbf{ForeSight overview.} ForeSight profiles the kernel once on \Czero, replays its CTAs on \Cone, and solves the model at each moment. The right panel enlarges one moment $t^*$.}
  \label{fig:foresight}
\end{figure*}

\subsection{Validating Per-Unit Values}
\label{sec:us-validation}

To validate the correctness of UnitSight's per-unit values, we compute their sum, average, maximum, and minimum and compare them with CUPTI's results on the same CounterData. On H20, we check about 44,000 SM and SMSP metrics, about 29,000 L1 cache metrics, and the metrics of L2 slices, DRAM channels, and other units. All are exactly equal.

\looseness=-1 Next, we check whether each value belongs to the right unit. We construct test kernels to set the activity of each unit, and compare UnitSight's per-unit values with this activity. For SMs, a test kernel gives different SMs different numbers of active cycles. Each CTA records its SM with \texttt{\%smid} and its cycles with \texttt{clock64}. Compared with these records, UnitSight's per-SM active cycles have a median error of 0.0\% and a maximum error of 0.2\%. The small deviation is expected because the in-kernel clock reads and the PM counter cover slightly different cycles. For L2 slices, a test kernel uses the L2 address hash of H20 to read only the cache lines of one slice, and we repeat it for every slice. In every run, UnitSight counts at least 99.99\% of the L2 sectors on the targeted slice and nearly none on the other slices.

\subsection{Case Study: FlashAttention-3 Decode}
\label{sec:us-case}

To show how UnitSight diagnoses a real LLM kernel, we profile vLLM's FlashAttention-3 (FA3) decode kernel. FA3 is a persistent kernel: each SM runs one CTA, with a producer warpgroup loading K and V and a consumer warpgroup running the MMAs. For a Llama-3.1-8B shape with batch size 16 and context length 8,192, the kernel takes 359 $\mu$s on H20. First, the per-SM values from range profiling show that 50 SMs execute about twice as many instructions as the other 28 SMs. This is because the kernel has 128 tiles for 78 SMs, so 50 SMs process two tiles and the other 28 SMs process only one. Therefore, we split each sequence's KV into three parts to balance the tiles across the SMs. This makes the kernel 11\% faster, reducing runtime from 359 $\mu$s to 319 $\mu$s.

Second, we apply in-kernel PM reading. For each KV block, the producer and the consumer read the PM registers and the clock at the start and end of each of their steps. The consumer waits for data only 6\% of its time, whereas the producer spends more than half its time waiting for the consumer to free buffer slots. The bottleneck is therefore the consumer's computation. The PM registers further show that the tensor cores reach about 88\% of H20's BF16 peak during the consumer's compute steps, so they limit the consumer. Since the tensor cores limit the kernel, we run the attention in FP8 instead of BF16, which the tensor cores compute twice as fast. This reduces the runtime from 359 $\mu$s to 223 $\mu$s, by 38\%.

For comparison, we also profile the kernel with NCU. NCU averages the stalls of all warps on all SMs, mixing the producer and the consumer. Its performance diagnosis ranks memory latency and occupancy as the top optimization opportunities, while the actual bottleneck is the tensor cores. Following this diagnosis, we optimize the memory side by enlarging the KV page size, but the runtime does not improve (362--364 $\mu$s). UnitSight finds the actual bottleneck because in-kernel PM reading measures each step of the producer and the consumer separately.

\section{ForeSight}
\label{sec:foresight}

This section presents \fs, a queueing analytical model for GPU performance prediction. From one profile of a kernel on the original GPU \Czero (Section~\ref{sec:fs-profile}, \step{1}), \fs predicts its execution time on a target configuration \Cone (Figure~\ref{fig:foresight}). It replays the CTAs of the kernel on \Cone (Section~\ref{sec:fs-replay}, \step{2}). At each moment of the replay, it solves a queueing model that reuses the fixed-latency stalls measured on \Czero and recomputes the queueing at each hardware unit on \Cone (Section~\ref{sec:fs-solve}, \step{3}). Section~\ref{sec:fs-params} lists the configurable parameters of the target configuration.

\subsection{Profiling on \Czero}\label{sec:fs-profile}

\fs starts by profiling the kernel on \Czero with three tools: \us, SASS analysis with NCU, and NVBit (Figure~\ref{fig:foresight}, \step{1}). \fs uses this profile to predict the kernel's execution time on every target configuration \Cone, without profiling the kernel again.

In this profile, \us gives the counter values of each SM, each L2 slice, and each DRAM channel. However, the units of \Cone can differ from those of \Czero, so these values cannot be used directly on \Cone. For the SMs, \Cone runs the same CTAs, but its number of SMs can differ, which changes how many CTAs each SM receives and thus the instructions it executes and its counter values. \fs therefore sums the counter values of all SMs on \Czero and divides them by the total number of instructions these SMs execute to obtain the average per instruction, and uses these averages for every SM of \Cone. For the slices and channels, the SMs send the same requests, but the number of these units on \Cone can differ, so the requests each unit receives differ as well. \fs therefore divides the counter values of each slice and each channel on \Czero by the total number of requests sent by the SMs, obtaining the average number of times a request reaches each unit. On GPUs with two L2 partitions, \fs computes the averages of the slices separately for the SMs of each partition, because each SM sends its requests only to the slices of its own partition (Section~\ref{sec:motivation}).

Range profiling in \us accumulates these values over the whole kernel. A kernel, however, executes different parts of its code at different times. \fs therefore divides the kernel into smaller regions of SASS code called \emph{phases} (phase~0 and phase~1 in Figure~\ref{fig:foresight}, \step{2}), using static analysis of the SASS code. A new phase begins after each CTA barrier, because all warps in a CTA synchronize at the barrier. Phase boundaries also fall before and after each outermost loop, because the loop repeats the same code and naturally forms one phase. To distribute these values across the phases, \fs profiles the kernel with the Source Counters section of NCU (NCU SASS). For every SASS instruction, NCU SASS reports how many times the instruction executes and how many memory requests it issues. \fs then splits the values in proportion to the executions and memory requests in each phase. The static SASS analysis also gives the pipeline on which each instruction executes and the dependences between instructions.

On \Cone, when the L2 capacity changes or a different set of SMs runs together, the same memory requests hit differently in the L1 and L2 caches, changing the counter values of the L2 slices and DRAM channels. To estimate this change, \fs instruments the kernel with NVBit and records the memory addresses accessed by each warp (Figure~\ref{fig:foresight}, \step{1}). From these addresses, \fs simulates the L1 and L2 caches on both \Czero and \Cone using reuse distance, and then applies the simulated change to the counter values measured by \us. For warp-specialized kernels (Section~\ref{sec:fs-solve}), NVBit also records the instructions of each warp and its synchronization instructions in execution order, and \fs uses this record to simulate the synchronization between warps.

\subsection{Schedule Replay}\label{sec:fs-replay}

On \Cone, each SM can receive different CTAs than on \Czero (Section~\ref{sec:fs-profile}), and the execution time of the kernel is determined by which CTAs each SM receives and when it runs them. \fs therefore replays the execution of the kernel on \Cone by simulating the GPU block scheduler and dispatching the kernel's CTAs to the SMs (Figure~\ref{fig:foresight}, \step{2}). Like the hardware, the block scheduler dispatches CTAs in index order, assigning each to the SM with the fewest resident CTAs. Each SM holds at most as many CTAs as occupancy allows. When a CTA ends, the next CTA in index order takes its place.

The execution time of a CTA depends on its work and on the number of instructions it executes per cycle (IPC). The work $W$ is the same for every CTA: the total number of instructions executed by all SMs on \Czero (Section~\ref{sec:fs-profile}) divided by the number of CTAs. The IPC of a CTA depends on the CTAs resident on its SM and on their phases. These CTAs share the pipelines and the memory path of the SM, and how heavily a CTA uses them depends on its phase. These IPCs therefore change only when a CTA starts, ends, or enters its next phase. We call such a moment an \emph{event} (the dashed lines in Figure~\ref{fig:foresight}). At each event, \fs computes the IPC of every CTA with the model of Section~\ref{sec:fs-solve} ($t^*$ in Figure~\ref{fig:foresight}). From these IPCs, \fs finds the moment at which each CTA on each SM would finish its current phase. The earliest of these moments is the next event, and \fs advances all CTAs to it. The replay continues in this way until the last CTA ends. The execution time of the kernel is this moment plus the fixed launch overhead.

In this replay, CTAs starting together on an SM have the same work and IPC, so they always enter each phase together and end together. On a real GPU, however, the latency of each memory access varies with cache hits and queueing, so these CTAs enter their phases and end at different moments. To model these differences, \fs makes the work of the CTAs follow a normal distribution before the replay. Its mean is $W$, and its standard deviation is $c_v W$. Here $c_v$ is the coefficient of variation of the CTA execution times, the ratio of their standard deviation to their mean. When each SM of \Czero runs a single CTA, the execution time of each SM is that of its CTA, so \fs takes $c_v$ directly as the coefficient of variation of the cycles per instruction of the SMs, measured by \us. When each SM runs several CTAs, the execution time of an SM no longer gives that of a single CTA, so \fs instead estimates $c_v$ from the variation of the memory latency:
\begin{equation}
c_v = \frac{\sqrt{V_w\,n\,\sigma^2}}{T},
\label{eq:cta-spread}
\end{equation}
where the numerator is the standard deviation of the execution time of a CTA and $T$ is its mean, computed by the model of Section~\ref{sec:fs-solve}. A warp waits for data from the L2 or DRAM $n$ times, as found by the SASS analysis of Section~\ref{sec:fs-profile}, and the memory latency of each wait has variance $\sigma^2$, also computed by the model of Section~\ref{sec:fs-solve}. The execution time of a warp therefore has variance $n\,\sigma^2$. A CTA ends with the slowest of its $w$ warps, so this variance is multiplied by $V_w$, the variance of the maximum of $w$ standard normal variables.

\subsection{Solving One Moment}\label{sec:fs-solve}

\noindent\textbf{SM model.} At each event $t^*$, \fs computes the IPC of every CTA on each SM (Figure~\ref{fig:foresight}, \step{3}). The IPC of a CTA is the sum of the IPCs of its warps, because the warp scheduler interleaves their instructions. The IPC of a warp is $1/\tau$, where the \emph{issue interval} $\tau$ is the average number of cycles from issuing one instruction to issuing the next. \fs breaks the issue interval down by the reason the warp stalls:
\begin{equation}
\tau = d^{\mathrm{fix}} + d^{\mathrm{SFU}} + \sum_{k} \tau_{k} + b + m.
\label{eq:inst-latency}
\end{equation}
The fixed-latency stall $d^{\mathrm{fix}}$ covers the stalls of a fixed number of cycles, such as waiting for the result of a pipeline or of a shared-memory access, or for the target of a branch. As Section~\ref{sec:motivation} shows, it does not change on \Cone, and \fs carries it from \Czero: $d^{\mathrm{fix}}$ equals the cycles of these stall reasons given by \us, averaged per instruction as in Section~\ref{sec:fs-profile}. The SFU stall $d^{\mathrm{SFU}}$ is also a fixed-latency stall and does not change on \Cone. It is the stall on the results of the SFU. However, \us has no counter for SFU stalls, so \fs computes $d^{\mathrm{SFU}}$ separately rather than including it in $d^{\mathrm{fix}}$. Using the SASS analysis, \fs finds each SFU instruction and the instruction depending on it. $d^{\mathrm{SFU}}$ equals the latency of the SFU, measured with a microbenchmark, minus the cycles of the other instructions between the two.

Because warps share the warp scheduler, the pipelines, and the L1 cache, they queue to use these units. At each unit $k$, the cycles a warp spends queueing there form the SM queueing stall $\tau_k$ ($\sum_k \tau_k$ in Eq.~\eqref{eq:inst-latency}), which depends on the \emph{processing time} and the \emph{queue length} of the unit. The processing time is the average number of cycles a unit spends per instruction and depends only on the instructions and the unit itself, not on the warps resident on the SM. The warp scheduler issues one instruction per cycle, so its processing time is always one cycle. For the pipelines and the L1 cache, the cycles spent processing instructions come from the counter values of \us, averaged per instruction as in Section~\ref{sec:fs-profile}. Since \us provides no such counter for the SFU, \fs instead multiplies the number of SFU instructions by the cycles the SFU needs to process one, measured with a microbenchmark. In contrast, the queue length of a unit is the average number of warps waiting there and changes with the warps resident on the SM, so \fs recomputes $\tau_k$ on \Cone. The issue interval $\tau$ also affects the queue length: a shorter $\tau$ makes each warp reach the unit more often and lengthens the queue. However, $\tau$ itself includes $\tau_k$ (Eq.~\eqref{eq:inst-latency}), so the queue length and $\tau_k$ depend on each other, and \fs solves them jointly with mean value analysis (MVA)~\cite{reiser1980mva}. MVA is a standard method of queueing theory: for a fixed number of warps sharing a set of units, it computes the mean queue length and queueing time $\tau_k$ at each unit. Starting from initial queue lengths, MVA computes $\tau_k$ in each iteration from the queue length and processing time of every unit, which yields $\tau$ and the total IPC of the warps sharing the units. Multiplying this total IPC by $\tau_k$ gives the new queue lengths, and MVA iterates until they converge. \fs runs MVA separately on each SMSP and obtains the IPC of the warps of each phase, which add up to the IPC of each CTA.

The barrier stall $b$ ($b$ in Eq.~\eqref{eq:inst-latency}) is the number of cycles a warp waits at a CTA barrier for the other warps of the CTA to arrive. The warps of a CTA arrive at a barrier at different times for two reasons: between two barriers, the warps may execute different numbers of instructions, and the latencies of their memory accesses vary. These differences in arrival time change on \Cone with the cycles per instruction and the queueing of memory requests, so \fs recomputes $b$. \fs treats the execution time $T_i$ of each warp $i$ between two barriers as normally distributed, with a variance from these two sources. The barrier stall is the amount by which the expected longest execution time exceeds the mean, averaged per instruction:
\begin{equation}
b = \beta \left( \mathrm{E}\!\left[\max_{1 \le i \le w} T_i\right] - \mathrm{E}[T_i] \right),
\label{eq:barrier}
\end{equation}
where $\beta$ is the number of barriers per instruction and $w$ is the number of warps of a CTA.

The memory queueing stall $m$ ($m$ in Eq.~\eqref{eq:inst-latency}) is the stall on data from the L2 or DRAM. It depends on the memory latency $R$, which is the average round-trip time of a request from an SM and changes with the traffic of all SMs. \fs therefore recomputes $m$ on \Cone, with the $R$ computed by the memory model below. After issuing a memory instruction, a warp continues with the instructions that do not depend on it, which \fs finds with the SASS analysis. The memory queueing stall equals $R$ minus the cycles the warp spends executing these instructions.

\textbf{Memory model.} The memory model computes the memory latency $R$ for each SM. A request from an SM leaves the L1 cache, travels through the crossbar to an L2 slice, and on an L2 miss reaches a DRAM channel (Figure~\ref{fig:foresight}, \step{3}). The memory latency of SM $s$ is
\begin{equation}
R_s = L_{0,s} + \sum_k v_{s,k}\, q_k,
\label{eq:latency}
\end{equation}
\looseness=-1 where $L_{0,s}$ is the latency of SM $s$ without queueing, $v_{s,k}$ is the average number of times a request of SM $s$ reaches memory unit $k$, and $q_k$ is the average number of cycles a request queues at unit $k$. \fs measures $L_{0,s}$ with a microbenchmark. It obtains $v_{s,k}$ from the counter values of \us and, on \Cone, adjusts it by the change in the request counts simulated with reuse distance (Section~\ref{sec:fs-profile}). The queueing delay $q_k$ grows with the number of requests unit $k$ receives per cycle, and microbenchmarks measure this queueing delay as a function of the requests per cycle for the L2 slices and the DRAM channels. Unit $k$ receives $\sum_s v_{s,k} \lambda_s$ requests per cycle, where $\lambda_s$ is the number of requests SM $s$ sends per cycle, so \fs obtains $q_k$ from this function at $\sum_s v_{s,k} \lambda_s$. The memory model also computes the variance $\sigma^2$ of the memory latency, from the different read paths of the requests and from the variation of the queueing delay. The barrier stall (Eq.~\eqref{eq:barrier}) and $c_v$ in Section~\ref{sec:fs-replay} (Eq.~\eqref{eq:cta-spread}) use this variance.

\textbf{Coupling the two models.} The two models depend on each other, because the SM model needs $R$, and the memory model needs $\lambda_s$, which follows from the IPCs computed by the SM model and the number of requests per instruction. \fs therefore solves the two models together with the fixed-point method~\cite{ciardo1993decomposition}. Starting from the $R$ solved at the previous event, the SM model computes the IPC of each CTA to give $\lambda_s$, and the memory model then computes a new $R$. The iteration repeats until the new $R$ equals the $R$ it started from. At each iteration, \fs chooses the next $R$ with Newton's method~\cite{kelley1995iterative}: from the derivative of the new $R$ with respect to $R$, it estimates the $R$ at which the new $R$ equals the starting one.

\begin{figure}[t]
  \centering
  \includegraphics[width=\columnwidth]{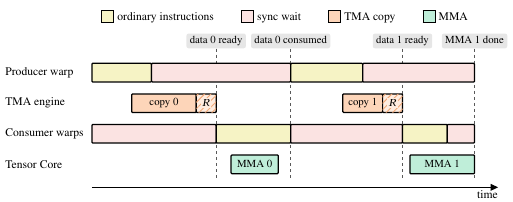}
  \caption{\textbf{Event simulation of a warp-specialized CTA.}}
  \label{fig:ws}
\end{figure}

\textbf{Warp-specialized kernels.} Recent GPUs such as Hopper provide a TMA engine and asynchronous MMA instructions on the Tensor Core, which allow a kernel to use warp specialization: the warps of a CTA execute different instructions and synchronize through barriers in shared memory (Figure~\ref{fig:ws}). To support such kernels, \fs uses NVBit on \Czero to record the instructions each warp executes and the synchronization relations between warps, neither of which changes with the configuration (Section~\ref{sec:fs-profile}). \fs then simulates the warps within a CTA on \Cone in time order, computing the execution time of each warp between two synchronizations and the time of each synchronization.

Warp-specialized kernels use two classes of instructions, which differ in how their dependences are enforced. Dependences on ordinary instructions are enforced automatically by stalling the issue of any instruction that uses a pending result. Explicit synchronization enforces dependences on asynchronous instructions (TMA copy instructions on the TMA engine and asynchronous MMA instructions on the Tensor Core), so a warp stalls for an asynchronous instruction only at a synchronization. The execution time of a warp between two synchronizations therefore depends only on its ordinary instructions. For those instructions, \fs keeps their execution time measured on \Czero and recomputes only their memory queueing stall with the memory latency $R$ on \Cone. We find that ordinary instructions in warp-specialized kernels barely queue inside the SM (0.9\% of warp execution time on the H20), so \fs does not recompute their SM queueing stall with MVA.

The time of a synchronization is the moment when all participating warps have reached their synchronization instructions and all asynchronous instructions it depends on have completed. \fs obtains when the warps reach their synchronization instructions from the execution times of their ordinary instructions. It obtains when the asynchronous instructions complete from their execution times, computed as the number of operations divided by the Tensor Core throughput for an MMA instruction and as the size divided by the TMA engine bandwidth plus the memory latency $R$ for a TMA copy instruction. From these two times, \fs computes the time of each synchronization in time order until all warps of the CTA finish. The completion time of the last warp gives the CTA execution time, by which the CTA progresses in the replay (Figure~\ref{fig:foresight}, \step{2}) instead of the IPC from the SM model.

\looseness=-1 Figure~\ref{fig:ws} shows this simulation on an example CTA: a producer warp loads data into shared memory with TMA copy instructions, consumer warps compute on these data on the Tensor Core, and the producer loads the next data into the same location only after they are consumed. In time order, \fs first computes the time of the producer's ordinary instructions (yellow). The producer issues copy~0 (orange) during them. Copy~0 completes after the memory latency $R$ (hatched) and triggers the consumers' first synchronization (data~0 ready), which ends the consumer stall (pink). Next, \fs computes the time of the consumers' ordinary instructions, during which they issue MMA~0 (green). The second synchronization depends on both these ordinary instructions and MMA~0. The ordinary instructions finish later, so the synchronization happens when they finish (data~0 consumed). The producer then resumes and issues copy~1. In the second round, MMA~1 finishes after the ordinary instructions and triggers the synchronization (MMA~1 done).

\subsection{Configurable Parameters}\label{sec:fs-params}

\fs allows users to change the parameters of the SMs, the L2, and DRAM starting from \Czero{} to obtain a target configuration \Cone{} and to predict the execution time of the kernel on \Cone. For the SMs, users can change the number of SMs, their distribution across the GPCs, and the SM clock. Inside each SM, the parameters are the L1 capacity, the throughput of each pipeline (ALU, FMA, FP64, tensor, TMA, and SFU in Figure~\ref{fig:foresight}), the latency of the SFU, and the limits of resources such as registers, shared memory, and warps. The L2 of \Cone{} may differ from that of \Czero{} in the number of slices, the capacity, and the clock. The latency without queueing of an L2 hit and the queueing delay of each slice as a function of the requests per cycle can also be changed. The DRAM bandwidth can be adjusted through the number of channels and the memory clock. As with the L2, the queueing delay of each channel and the latency without queueing of a DRAM access are configurable.

\section{Evaluation}
\label{sec:eval}

This section evaluates how accurately and how fast \fs predicts the performance of kernels and of end-to-end LLM inference on target configurations.
To check these predictions, we construct the target configurations on real GPUs and measure the execution time of each kernel (Section~\ref{sec:eval-setup}).
We compare the accuracy of \fs and existing methods by the mean absolute percentage error (MAPE) of their predictions against the measured times (Section~\ref{sec:eval-accuracy}).
Finally, we compare their prediction speed (Section~\ref{sec:eval-speed}).

\subsection{Experimental Setup}\label{sec:eval-setup}

\begin{table}[t]
\caption{\textbf{GPUs in the evaluation.} The numbers of SMs, L2 slices, and DRAM channels are observed by \us.}
\label{tab:setup}
\small\setlength{\tabcolsep}{1pt}
\begin{tabular*}{\columnwidth}{@{\extracolsep{\fill}}llcccccc@{}}
\toprule
\raisebox{-.5\normalbaselineskip}[0pt][0pt]{GPU} & \raisebox{-.5\normalbaselineskip}[0pt][0pt]{Arch.} & \raisebox{-.5\normalbaselineskip}[0pt][0pt]{SMs} & \multicolumn{2}{c}{L2} & \multicolumn{3}{c}{DRAM} \\
\cmidrule(lr){4-5}\cmidrule(l){6-8}
 & & & MB & slices & type & channels & TB/s \\
\midrule
A100 SXM      & Ampere    & 108 & 40  & 80 & HBM2e   & 40 & 2.04 \\
RTX 6000 Ada  & Ada       & 142 & 96  & 48 & GDDR6   & 12 & 0.96 \\
RTX 4090      & Ada       & 128 & 72  & 36 & GDDR6X  & 12 & 1.01 \\
H20          & Hopper    & 78  & 60  & 96 & HBM3    & 48 & 4.00 \\
H100 SXM      & Hopper    & 132 & 50  & 80 & HBM3    & 40 & 3.35 \\
H100 PCIe     & Hopper    & 114 & 50  & 80 & HBM2e   & 40 & 2.00 \\
DGX Spark     & Blackwell & 48  & 24  & 16 & LPDDR5x & 16 & 0.27 \\
RTX PRO 6000  & Blackwell & 188 & 128 & 64 & GDDR7   & 16 & 1.60 \\
RTX 5090      & Blackwell & 170 & 96  & 48 & GDDR7   & 16 & 1.79 \\
\bottomrule
\end{tabular*}
\end{table}

\noindent\textbf{Platforms and workloads.}
We evaluate \fs on nine GPUs of four architectures (Table~\ref{tab:setup}).
On these GPUs, \fs predicts the execution time of 50 workloads on each target configuration.
Forty-seven of these workloads are operators executed by vLLM~\cite{kwon2023vllm} during LLM inference, including GEMM, attention, MoE, normalization, activation, rotary embedding, and quantization.
Their input shapes follow the Llama-3.1 (8B, 70B)~\cite{grattafiori2024llama3} and Qwen3 (8B, 32B, 30B-A3B)~\cite{yang2025qwen3} models.
On Hopper and Blackwell, vLLM selects warp-specialized kernels (Section~\ref{sec:fs-solve}) for part of the attention and GEMM operators.
On Hopper, for example, 24 of the vLLM kernels are warp-specialized.
The other three workloads are microbenchmarks testing whether \fs accurately predicts kernels dominated by DRAM reads and writes or by FP32 FMA computation.

\textbf{Target configurations.}
To evaluate \fs when the number of SMs and their distribution across GPCs change, we construct target configurations with green contexts~\cite{nvidia2026greenctx}, which restrict the SMs available to a kernel.
For clock changes, the target configurations are the original GPU with a locked SM frequency.
For memory-side changes together with the SMs, we use MIG instances~\cite{nvidia2026mig} as target configurations, each with fewer SMs, L2 slices, and DRAM channels and a smaller L2.
Finally, to evaluate predictions across GPUs, we collect the profile on one Ada, one Hopper, and one Blackwell GPU and use the other GPUs of the same architecture as target configurations.
These GPUs differ from the GPU of \Czero in the number of SMs, the L2 capacity, and the DRAM bandwidth (Table~\ref{tab:setup}).
The original GPU without any of these changes is also a target configuration.
On some MIG instances and on another GPU, vLLM selects a different kernel or grid for a workload, and we evaluate only the workloads whose kernel and grid stay the same.
We run each kernel 30 times on each target configuration and take the median as the measured execution time.
Together, these configurations cover SM, clock, and memory-side changes within a GPU and changes across GPUs, all on real GPUs.
Prior evaluations on real GPUs typically test one type of change: prediction across GPUs~\cite{yu2021habitat,lee2025neusight,zhang2026pipeweave}, changes of the SM count only~\cite{bullet2025}, or changes of the frequency only~\cite{wang2018gpgpu}.
Other studies evaluate several types of parameter changes, but take simulator results as the ground truth~\cite{lee2022gcom,wang2020mdm,cha2025gcscaler}.

\begingroup\emergencystretch=1em%
\textbf{Baselines.}
We select three classes of existing methods as baselines and run each as described in its paper: (i) the simulator Accel-Sim~\cite{khairy2020accelsim}, (ii) the analytical models DuetServe~\cite{duetserve2025}, Bullet~\cite{bullet2025}, LLMCompass~\cite{zhang2024llmcompass}, Habitat~\cite{yu2021habitat}, Wang and Chu~\cite{wang2018gpgpu}, GCoM~\cite{lee2022gcom}, AMALI~\cite{cao2025amali}, and a roofline model computed from datasheet peaks~\cite{williams2009roofline}, and (iii) the data-driven models NeuSight~\cite{lee2025neusight} and PipeWeave~\cite{zhang2026pipeweave}.
Because Habitat's MLPs support only FP32 matrix multiplications and ours are BF16 or FP8, we use only its wave scaling.
\looseness=-1 Some baselines do not support all workloads or target configurations: NeuSight, PipeWeave, and LLMCompass support only a subset of operators, Accel-Sim, GCoM, and AMALI do not support Blackwell GPUs, and Wang and Chu supports only frequency changes.
We compare each baseline with \fs only on the workloads and target configurations supported by that baseline.\par\endgroup

\subsection{Prediction Accuracy}\label{sec:eval-accuracy}

\begin{table}[t]
\caption{\textbf{MAPE (\%) of the baselines and \fs on each GPU.} All: all GPUs combined.}
\label{tab:accuracy}
\footnotesize\setlength{\tabcolsep}{1pt}
\begin{tabularx}{\columnwidth}{@{}l*{10}{>{\raggedleft\arraybackslash}X}@{}}
\toprule
Method & \makecell[r]{A100\\SXM} & \makecell[r]{6000\\Ada} & \makecell[r]{RTX\\4090} & H20 & \makecell[r]{H100\\SXM} & \makecell[r]{H100\\PCIe} & \makecell[r]{DGX\\Spark} & \makecell[r]{PRO\\6000} & \makecell[r]{RTX\\5090} & All \\
\midrule
Accel-Sim & 103 & 25.8 & 30.4 & 20.4 & 16.3 & 18.1 & -- & -- & -- & 32.3 \\
\midrule
DuetServe & 32.6 & 28.9 & 24.5 & 35.4 & 26.5 & 26.8 & 15.5 & 28.3 & 29.7 & 28.1 \\
Bullet & 32.7 & 45.0 & 24.3 & 35.9 & 27.0 & 27.6 & 15.1 & 28.3 & 29.9 & 29.7 \\
LLMCompass & 128 & 157 & 125 & 58.3 & 134 & 149 & 159 & 138 & 114 & 129 \\
Habitat & 14.9 & 11.4 & 9.3 & 18.6 & 16.3 & 15.6 & 9.9 & 12.0 & 13.7 & 14.1 \\
Wang \& Chu & 99.9 & 95.0 & 65.4 & 45.9 & 99.3 & 93.7 & 110 & 129 & 108 & 94.0 \\
GCoM & 96.1 & 58.4 & 67.6 & 89.6 & 71.5 & 62.1 & -- & -- & -- & 73.8 \\
AMALI & 28.5 & 44.9 & 45.2 & 56.1 & 51.2 & 49.5 & -- & -- & -- & 47.0 \\
Roofline & 42.2 & 28.7 & 25.7 & 46.5 & 37.2 & 33.6 & 20.7 & 32.7 & 31.5 & 34.3 \\
\midrule
NeuSight & 292 & 379 & 241 & 151 & 362 & 413 & 279 & 363 & 224 & 311 \\
PipeWeave & 23.1 & 29.5 & 28.8 & 51.2 & 24.4 & 30.2 & 43.8 & 26.4 & 35.9 & 31.7 \\
\midrule
\rowcolor{oursbg}\fs (ours) & 10.1 & 9.0 & 10.1 & 4.7 & 7.8 & 8.0 & 5.8 & 8.3 & 11.9 & 8.3 \\
\bottomrule
\end{tabularx}
\end{table}

\noindent\textbf{Overall accuracy.}
Table~\ref{tab:accuracy} reports the MAPE of \fs and each baseline on each GPU. The MAPE of \fs ranges from 4.7\% on H20 to 11.9\% on RTX 5090 across individual GPUs, and is 8.3\% across all GPUs combined. Every baseline has a higher MAPE than \fs on every GPU (9.9\% to 413\%), except Habitat on RTX 4090 (9.3\% vs.\ 10.1\%).

\begin{figure*}[t]
\centering
\includegraphics[width=\textwidth]{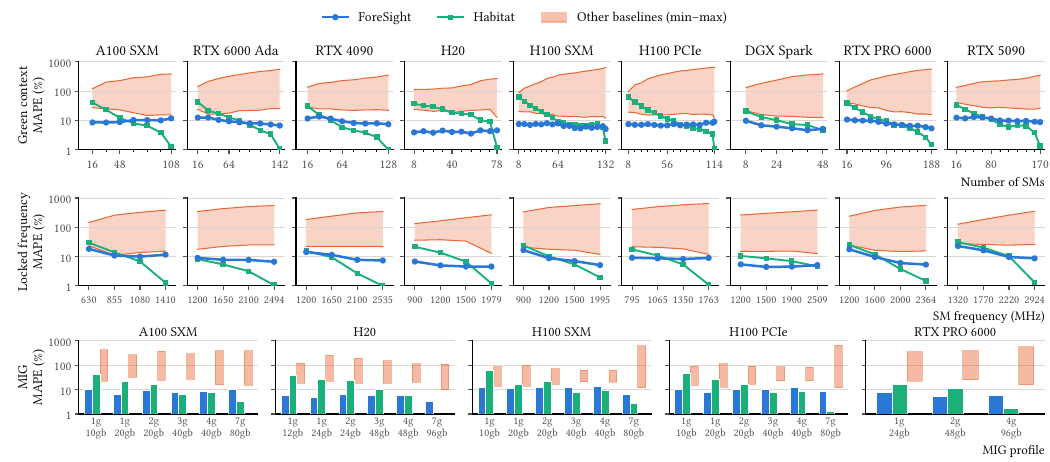}
\caption{\textbf{MAPE of \fs and the baselines on each target configuration.} Each column is a GPU. The rows vary the number of SMs with green contexts, the SM frequency, and the MIG profile. The rightmost point of the first two rows is the original GPU. At each configuration, the band (the bars in the MIG row) spans the lowest to the highest MAPE of the other baselines.}
\label{fig:acc}
\end{figure*}

\looseness=-1 \textbf{Accuracy on each configuration.}
Figure~\ref{fig:acc} shows the MAPE on each target configuration.
On every target configuration, \fs has a lower MAPE than almost all baselines, and on each GPU its MAPE varies little across configurations while those of the baselines vary widely.
Habitat is the only baseline with a lower MAPE than \fs on some configurations.
It uses wave scaling to scale a kernel's measured execution time on the original GPU to the target configuration, so it is accurate on configurations near the original GPU.
However, its error grows as the target configuration moves further from the original GPU, and exceeds that of \fs on most configurations.
\fs also has a much lower worst-case MAPE than every baseline: its MAPE does not exceed 23.1\% on any target configuration, whereas every baseline exceeds 43\% on its own worst one.

\looseness=-1 \textbf{Across SM counts.}
When the number of SMs changes, \fs stays accurate on every GPU, whereas the errors of the baselines change widely with the SM count (Figure~\ref{fig:acc}, green context).
Using green contexts, we vary the number of SMs from 8 with a stride of 8 on the Hopper GPUs and DGX Spark, and from 16 with a stride of 16 on the other GPUs.
Across these SM counts, \fs's MAPE ranges from 3.5\% to 13.6\% and varies by at most 6.2 percentage points on each GPU, whereas every baseline varies more.
On RTX 6000 Ada, for example, from 16 SMs to the original GPU, the MAPE of AMALI rises from 29\% to 59\%, whereas that of Habitat drops from 42\% to 1.1\%.

\textbf{Across SM frequencies.}
When the SM frequency changes, \fs remains more accurate than almost all baselines, including Wang and Chu, which specifically models frequency changes (Figure~\ref{fig:acc}, locked frequency).
We use \texttt{nvidia-smi} to lock the SM frequency of each GPU at three lower levels, the lowest at roughly half the default frequency.
On H20, for example, we lock it at 900, 1200, and 1500~MHz, whereas the original GPU runs at 1979~MHz.
At these frequencies, \fs's MAPE ranges from 4.4\% to 23.1\%, whereas that of Wang and Chu ranges from 46\% to 130\%.

\looseness=-1 \textbf{Across MIG instances.}
When the memory side changes together with the SMs, \fs also stays accurate, whereas the errors of the baselines vary widely across MIG profiles (Figure~\ref{fig:acc}, MIG).
On the five GPUs that support MIG, we evaluate every instance size (MIG profile), such as \texttt{1g.10gb}, which has one compute slice and 10~GB of memory together with the corresponding L2 slices and DRAM channels.
On these profiles, \fs's MAPE ranges from 3.4\% to 14.0\% and varies by at most 7.6 percentage points on each GPU, whereas every baseline varies more.
On A100 SXM, for instance, from \texttt{1g.10gb} to \texttt{7g.80gb}, the MAPE of Accel-Sim rises from 74\% to 141\%, whereas that of Habitat drops from 42\% to 3.3\%.

\textbf{Across GPUs.}
When \Cone is another GPU of the same architecture, \fs also stays more accurate than every baseline.
We test five pairs of a source GPU (\Czero) and a target GPU (\Cone): RTX 6000 Ada to RTX 4090, H20 to H100 SXM and H100 PCIe, and RTX PRO 6000 to RTX 5090 and DGX Spark.
Over these pairs, \fs's MAPE ranges from 7.5\% to 12.9\%, and on each pair it is lower than that of every baseline.
The closest baseline is Habitat from RTX 6000 Ada to RTX 4090, with a MAPE of 9.4\% against 7.6\% for \fs, but its MAPE reaches 47.4\% from H20 to H100 PCIe.

\looseness=-1 \textbf{End-to-end accuracy.}
To evaluate the prediction accuracy of \fs in real end-to-end LLM inference, we deploy Llama-3.1-8B with vLLM on an H20 GPU to process 64 requests, each with 2048 input and 2048 output tokens.
The target configurations are the original H20 GPU and green contexts with 8 to 72 SMs at a stride of 8.
vLLM processes these requests in iterations, each executing one forward pass of the model on the current batch, so \fs and the baselines predict the execution time of every kernel in every iteration and sum these predictions as the predicted total execution time.
We compare it with the end-to-end time measured on the same target configuration, from submitting the 64 requests to generating the last token.
For operators that a baseline does not support, we use the prediction of the roofline model.
Across all target configurations, \fs has a MAPE of 2.0\% and a maximum error of 3.0\%, whereas the MAPE of the baselines ranges from 14.3\% (AMALI) to 137\% (NeuSight).
For most baselines, most of the error comes from attention kernels, which account for 43\% to 52\% of the total execution time.
The baselines predict $0.06\times$ to $5.7\times$ the measured execution time of attention, whereas \fs predicts $1.02\times$ to $1.03\times$.

\looseness=-1 \textbf{Accuracy of per-unit predictions.}
To explain the accuracy of \fs, we use \us to measure the counter values of each SM, each L2 slice, and each DRAM channel on \Cone and check the per-unit predictions of \fs. On H100 PCIe, using green contexts with 16, 40, and 64 SMs, we run six kernels: copy, RMSNorm, SiLU, RoPE, FP8 quantization, and MoE. In Eq.~\eqref{eq:inst-latency}, the SM model breaks the issue interval $\tau$ down by the reason the warp stalls, and we use the counter values of \us to compute the fraction of $\tau$ taken by each stall. Averaged over the six kernels, each predicted fraction is within 6.2 percentage points of the measured fraction on every target configuration. For example, the memory queueing stall $m$ accounts for 41.9\% of $\tau$ in the prediction and 44.7\% in the measurement, and the fixed-latency stall $d^{\mathrm{fix}}$ for 15.5\% and 15.6\%. The memory model computes the memory latency $R$ from the requests received by each L2 slice and DRAM channel. We measure the requests at these units with \us, and \fs predicts them with a traffic-weighted error of 1\% to 17\%. For SiLU at 40 SMs, the imbalance across L2 slices shown in Section~\ref{sec:motivation} ($3.98\times$) is predicted as $3.87\times$.

\subsection{Prediction Speed}\label{sec:eval-speed}
\fs is faster than Accel-Sim and the interval-based models in both of its parts: collecting profiles on \Czero and predicting on each \Cone.
Like the instruction trace that Accel-Sim, GCoM, and AMALI take as input, the profile of \fs is collected on \Czero only once for each workload and reused for every target configuration.
On H20, \fs collects its profile faster than Accel-Sim's tracer collects the trace, with a median of 170~s against 233~s per workload.
On each \Cone, \fs replays the CTAs with its queueing model, whereas Accel-Sim simulates the trace cycle by cycle, and GCoM and AMALI analyze the trace of every warp.
\fs is faster than all three, with a median of 26~s on one CPU core to predict a workload on one \Cone, against 1272~s for Accel-Sim and 40~s and 108~s for GCoM and AMALI.
The other analytical and data-driven baselines are faster than \fs, with median times of 0.002~ms to 4.4~s, because most of them use formulas or trained models to compute execution time directly from workload features and hardware parameters without modeling workload execution on \Cone.
However, for the same reason, they cannot fully capture how workload execution changes with the target configuration, so their overall MAPE ranges from 14.1\% to 311\%, against 8.3\% for \fs (Table~\ref{tab:accuracy}).

\section{Case Study}
\label{sec:cases}

To show how \fs helps architects with DSE, we present a case study: we change the configuration of H20 to find the one with the highest performance/cost for 8B-model inference on a single GPU.
H20 has 78 SMs, a 60\,MB L2, and 6 HBM stacks, each providing 16\,GB of memory and about 0.67\,TB/s of bandwidth.
Each configuration keeps the SM design and clock frequency of H20 and changes only the number of SMs, the L2 capacity, and the number of HBM stacks.
The inference workload is the same as in the end-to-end evaluation of Section~\ref{sec:eval-accuracy}: Llama-3.1-8B~\cite{grattafiori2024llama3} in BF16 serving 64 requests, each with 2048 input tokens and 2048 output tokens, so each configuration needs at least 50\,GB of memory for the model weights and the KV cache.
From the profile of this workload on H20, \fs predicts the throughput of each configuration in output tokens per second, which we take as its performance.
We compute the cost of each configuration with the cost model of LLMCompass~\cite{zhang2024llmcompass}.
The cost is the sum of the on-chip die cost, which depends on the total area of the units, and the off-chip HBM cost, which depends on the number of HBM stacks.
We search 180 configurations, with 32 to 128 SMs, 10 to 60\,MB of L2, and 4 to 6 HBM stacks, to find the configuration with the highest performance/cost.

\begin{figure}[t]
  \centering
  \includegraphics[width=\columnwidth]{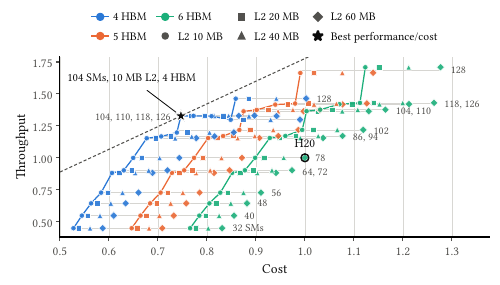}
  \caption{Predicted throughput and cost of every configuration, relative to H20.}
  \label{fig:case}
\end{figure}

Figure~\ref{fig:case} shows the predicted throughput and the cost of the 180 configurations, both normalized to H20.
Each gray line connects the configurations with the same number of SMs, and each colored line connects the configurations with the same number of HBM stacks and a 10\,MB L2.
More SMs give higher throughput, but the gain from added SMs is uneven: going from 64 to 72 SMs or from 86 to 94 SMs raises throughput by only 1\% to 2\%, going from 102 to 104 SMs raises it by 12\%, going further to 126 SMs adds only 5\%, and going from 126 to 128 SMs raises it by 20\%.
L2 capacity barely affects throughput: across all configurations, reducing the L2 from 60\,MB to 10\,MB changes throughput by at most 0.1\% and lowers cost by 8\% to 14\%.
HBM stacks also have little effect on throughput: with at most 102 SMs, reducing the HBM stacks from 6 to 4 lowers throughput by less than 2\% and cost by 24\% to 31\%.
With 104 or more SMs, however, the bandwidth of 4 HBM stacks starts to limit throughput, which becomes 2.7\% to 14\% lower than with 6 HBM stacks.
The performance/cost of a configuration equals the slope of the line from the origin to it, and the dashed line in Figure~\ref{fig:case} is the steepest of these lines and thus marks the highest performance/cost.
Therefore, the configuration we search for has 4 HBM stacks, a 10\,MB L2, and 104 SMs, and reaches 1.78$\times$ the performance/cost of H20, with 1.33$\times$ the throughput at 0.747$\times$ the cost.

\section{Related Work}
\label{sec:related}

\noindent\textbf{Simulation.}
Cycle-level simulators such as Accel-Sim~\cite{khairy2020accelsim} simulate each unit of a GPU cycle by cycle and can predict performance under any configuration.
However, they are slow.
To speed up simulation, sampling methods simulate only representative kernels~\cite{baddouh2021pka,chung2025stem}, warps and basic blocks~\cite{liu2023photon}, or thread block sets~\cite{yang2024hyfiss}, and scale-model simulation extrapolates from a GPU with fewer SMs~\cite{seyyedaghaei2024scalemodel}.
Unlike these works, \fs does not simulate the execution and takes tens of seconds per target configuration.

\textbf{Analytical models.}
Interval-based models compute the execution time from the issue and stall cycles of a representative warp on the target configuration.
GPUMech~\cite{huang2014gpumech} and GCoM~\cite{lee2022gcom} count the stalls of each type, GCScaler~\cite{cha2025gcscaler} scales the stall cycles of a simulated GPU, MDM~\cite{wang2020mdm} models the NoC and DRAM queueing caused by memory divergence, and AMALI~\cite{cao2025amali} adds tensor core and instruction cache models for LLM inference.
LLMCompass~\cite{zhang2024llmcompass} and TileSight~\cite{mo2026tilesight} estimate the execution time tile by tile.
Scaling-based models, closest to \fs, start from the execution time on the original GPU and scale it to the target configuration.
CRISP~\cite{nath2015crisp} scales it to other frequencies, Habitat~\cite{yu2021habitat} to other GPUs with wave scaling, and Van Lanker et al.~\cite{vanlanker2024predicting} to a new GPU with a hierarchical roofline model.
Unlike them, \fs reuses the fixed-latency stalls measured on the original GPU and recomputes the queueing at each hardware unit on the target configuration.
Earlier GPU queueing models use a single DRAM queue~\cite{wang2018gpgpu} or explain latency hiding~\cite{volkov2016latency} and shared-memory atomic bottlenecks~\cite{dong2025atomics}.

\looseness=-1 \textbf{Data-driven models.}
Wu et al.~\cite{wu2015gpgpu} pick a measured scaling curve with a neural network, NeuSight~\cite{lee2025neusight} predicts each tile with MLPs, and PipeWeave~\cite{zhang2026pipeweave} feeds per-pipeline demands from an analytical model to a machine learning model.
They describe a GPU only by its specifications, so they cannot capture how a workload loads each hardware unit.
Unlike these models, \fs needs no training and profiles each kernel on one GPU.

\textbf{GPU profilers.}
Profilers based on hardware counters, such as NCU~\cite{ncu} and CUPTI~\cite{cupti}, compute metrics from the PM counters but report only values aggregated across units.
Instrumentation tools such as Neutrino~\cite{huang2025neutrino} and KPerfIR~\cite{guan2025kperfir} record timestamps and program values but do not read the hardware PM counters.
Unlike these tools, \us recovers the counter values of each unit and lets a kernel read performance counters during execution.

\section{Conclusion}
\label{sec:conclusion}

\looseness=-1 This paper presents \fs, a queueing analytical model that predicts GPU performance across hardware configurations.
It builds on our observation that fixed-latency stalls stay the same across configurations, whereas queueing stalls vary with the pressure on each hardware unit.
\fs therefore reuses the fixed-latency stalls measured on GPU hardware and recomputes the queueing stalls of each unit from its pressure.
\us provides \fs with the per-unit counter values that NVIDIA's profilers hide.
On configurations of nine NVIDIA GPUs, \fs predicts kernel execution time with a MAPE of 8.3\%, compared with 14.1\% for the best baseline.
It also predicts the end-to-end execution time of LLM inference on vLLM with a MAPE of 2.0\%.
It runs faster than cycle-level simulators and interval-based models.
In a DSE case study, \fs finds an H20-derived configuration with 1.78$\times$ its performance/cost.

\bibliographystyle{ACM-Reference-Format}
\bibliography{refs}

\end{document}